\documentclass[10pt]{article}
\usepackage{epsf}
\usepackage{graphicx}

\begin{document}
\title
{How to verify the redshift mechanism of low-energy quantum
gravity}
\author
{Michael A. Ivanov \\
Physics Dept.,\\
Belarus State University of Informatics and Radioelectronics, \\
6 P. Brovka Street,  BY 220027, Minsk, Republic of Belarus.\\
E-mail:  michai@mail.by.}

\maketitle

\begin{abstract} In the model of low-energy quantum gravity by the
author,  the redshift mechanism is quantum and local, and it is
not connected with any expansion of the Universe. A few
possibilities to verify its predictions are considered here: the
specialized ground-based laser experiment; a deceleration of
massive bodies and the Pioneer anomaly; a non-universal character
of the Hubble diagram for soft and hard radiations; galaxy/quasar
number counts.
\end{abstract}
\section[1]{Introduction }
Many people consider the discovery of dark energy to be the main
finding of present cosmology. They are sure that an existence of
dark energy has been proved with observations of new, precise, era
of cosmology, and it is necessary only to clarify what it adds up.
Because of this, new cosmological centers are created and addicted
to this main goal. It seems to me that a new scientific myth has
risen in our eye; it is nice, almost commonly accepted, with
global consequences for physics, but it is really based on
nothing. What was a base for its rising? In 1998, two teams of
astrophysicists reported about dimming remote SN 1a \cite{2,3};
the one cannot be explained in the standard cosmological model on
a basis of the Doppler effect if the universe expands with
deceleration. Their conclusion that the Universe expands with
acceleration since some cosmological time served a base to
endenizen dark energy. But this conclusion is not a single
possible one; if the model does not fit observations, probably,
the one may simply be wrong.
\par If we stay on such the alternative point of view, what should
namely be doubt in the standard cosmological model? I think that
it should be at first its main postulate: a red shift is caused
with an expansion of the Universe. If this postulate is wrong,
then the whole construction of the model will wreck: neither the
Big Bang nor inflation, nor a temp or character of expansion would
not be interested. In the model of low-energy quantum gravity by
the author \cite{500},  the alternative redshift mechanism is
quantum and local. I review here a few possibilities to verify its
predictions.
\section[2]{Possibilities to verify the alternative redshift mechanism}
In my model \cite{500}, any massive body must experience a
constant deceleration $w \simeq - Hc$, where $H$ is the Hubble
constant and $c$ is the light velocity, of the same order of
magnitude as observed for NASA deep-space probes Pioneer 10/11
(the Pioneer anomaly) \cite{101,102}. This effect is an analogue
of cosmological redshifts in the model. Their common nature is
forehead collisions with gravitons. If my conjecture about the
quantum nature of this acceleration is true then an observed value
of the projection of the probe's acceleration on the sunward
direction $w_{s}$ should depend on accelerations of the probe, the
Earth and the Sun relative to the graviton background. It would be
very important to confront the considered model with observations
for small distances when Pioneer 11 executed its planetary
encounters with Jupiter and Saturn. In this period, {\it the
projection of anomalous acceleration may change its sign}
\cite{89}.
\par How to verify the main conjecture of this approach about the
quantum gravitational nature of redshifts in a ground-based laser
experiment? If the temperature of the background is $T=2.7 K$, the
tiny satellite of main laser line of frequency $\nu$ after passing
the delay line will be red-shifted at $\sim 10^{-3}$ eV/h and its
position will be fixed \cite{112}. It will be caused by the fact
that on a very small way in the delay line only a small part of
photons may collide with gravitons of the background. The rest of
them will have unchanged energies. The center-of-mass of laser
radiation spectrum should be shifted proportionally to a photon
path $l$. Then due to the quantum nature of shifting process, the
ratio of satellite's intensity to main line's intensity should
have the order: $\sim (h\nu / \bar{\epsilon})(H/ c) l,$ where
$\bar{\epsilon}$ is an average graviton energy. An instability of
a laser of a power $P$ should be only $\ll 10^{-3}$ if a photon
energy is of $\sim 1~eV$. It will be necessary to compare
intensities of the red-shifted satellite at the very beginning of
the path $l$ and after it. Given a very low signal-to-noise ratio,
one could use a single photon counter to measure the intensities.
When $q$ is a quantum output of a cathode of a used
photomultiplier, $N_{n}$ is a frequency of its noise pulses, and
$n$ is a desired ratio of a signal to noise's standard deviation,
then an evaluated time duration $t$ of data acquisition would have
the order: $ t= (\bar{\epsilon}^{2}c^{2} / H^{2}) (n^{2}N_{n} /
q^{2} P^{2} l^{2} ).$ Assuming $n=10,~N_{n}=10^{3}~s^{-1},~ q=0.3,
~P=100~ mW,~ l=100 ~m, $ we would have the estimate: $t= 200,000 $
years, that is unacceptable. But given $P=300~W$, we get: $t \sim
8$ days, that is acceptable for the experiment of such the
potential importance. Of course, one will rather choose a bigger
value of $l$ by a small laser power forcing a laser beam to
whipsaw many times between mirrors in a delay line - it is a
challenge for experimentalists. Maybe, it will be more convenient
to work with high-energy gamma rays to search for this effect in a
manner similar to the famous Pound-Rebka experiment \cite{113}.
\par The luminosity distance in this model is \cite{500}:
$D_{L}=a^{-1} \ln(1+z)\cdot (1+z)^{(1+b)/2},$ where $a=H/c,$ $z$
is a redshift. The theoretical value of relaxation factor $b$ has
been found in the assumption that in any case of a non-forehead
collision of a graviton with a photon, the latter leaves a photon
flux detected by a remote observer: $b=2.137$. It is obvious that
this assumption should be valid for a soft radiation when a photon
deflection angle is big enough and collisions are rare. It is easy
to find a value of the factor $b$ in another marginal case - for a
very hard radiation. Due to very small ratios of graviton to
photon momenta, photon deflection angles will be small, but
collisions will be frequent because the cross-section of
interaction is a bilinear function of graviton and photon energies
in this model. It means that in this limit case $b \rightarrow 0.$
For an arbitrary source spectrum, a value of the factor $b$ should
be still computed, and it will not be a simple task. It is clear
that $0 \leq b \leq 2.137,$ and in a general case it should depend
on a rest-frame spectrum and on a redshift. It is important that
the Hubble diagram is a multivalued function of a redshift: {\it
for a given $z,$ $b$ may have different values.} Theoretical
distance moduli $\mu_{0}(z) = 5 \log D_{L} + 25$ are shown in Fig.
1 for $b=2.137$ (solid), $b=1$ (dot) and $b=0$ (dash) \cite{114}.
If this model is true, all observations should lie in the
\begin{figure}[th]
\epsfxsize=8.0cm \centerline{\epsfbox{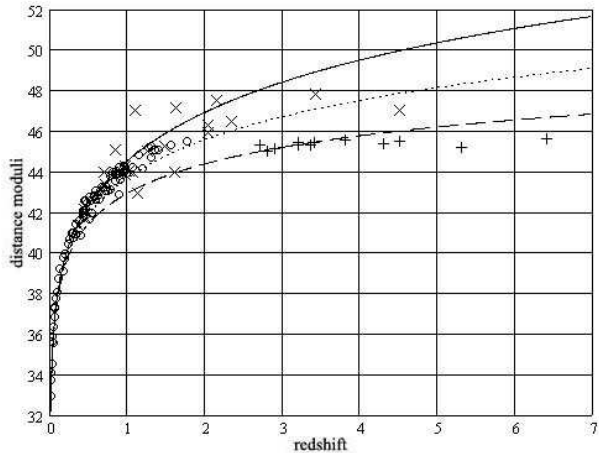}} \caption{
Hubble diagrams $\mu_{0}(z)$ with $b=2.137$ (solid), $b=1$ (dot)
and $b=0$ (dash); supernova observational data (circles, 82
points) are taken from Table 5 of \cite{203}, gamma-ray burst
observations are taken from \cite{204} (x, 24 points) and from
\cite{205} (+, 12 points for $z>2.6$).}
\end{figure}
stripe between lower and upper curves. For Fig. 1, supernova
observational data (circles, 82 points) are taken from Table 5 of
\cite{203}, gamma-ray burst observations are taken from \cite{204}
(x, 24 points) and from \cite{205} (+, 12 points for $z>2.6$). As
it was recently shown by Cuesta et al. \cite{4}, the Hubble
diagram with $b=1$ (in the language of this paper) gives the best
fit to the full sets of gamma-ray burst observations of
\cite{204,205} and it takes place in the standard FLRW cosmology
plus the strong energy condition. Twelve observational points of
\cite{205} belong to the range $z>2.6$, and one can see that these
points peak up the curve with $b=0$ which corresponds in this
model to the case of very hard radiation in the non-expanding
Universe with a flat space. In a frame of models without
expansion, any red-shifted source may not be brighter than it is
described with this curve.
\par In this model, the galaxy number counts/magnitude relation is
\cite{90}: $f_{3}(m)=(\phi_{\ast}\kappa / a^3)\cdot m\cdot
\int_{0}^{z_{max}} l^{\alpha+1}(m,z)\cdot \exp(-l(m,z))\cdot
(ln^{2}(1+z) / (1+z))dz.$  To compare this function with
observations by Yasuda et al. \cite{77}, let us choose the
normalizing factor from the condition: $f_{3}(16)=a(16), $ where $
a(m)\equiv A_{\lambda}\cdot 10^{0.6(m-16)} $ is the function
assuming "Euclidean" geometry and giving the best fit to
observations \cite{77}, $A_{\lambda}=const$ depends on the
spectral band; an upper limit is $z_{max} =10.$ In this case, we
have two free parameters - $\alpha$ and $L_{\ast}$ - to fit
observations, and the latter one is connected with a constant
$A_{1}\equiv A / a^{2}L_{\ast}$ if $l(m,z)=A_{1}f_{1}^{2}(z) /
\kappa^{m}.$ We have for $A_{1}$ by $H=2.14\cdot 10^{-18}\ s^{-1}$
(it is a theoretical estimate of $H$ in this model \cite{500}):
$A_{1}\simeq 5\cdot 10^{17}\cdot (L_{\odot} / L_{\ast}),$ where
$L_{\odot}$ is the Sun luminosity. Matching values of $\alpha$
shows that $f_{3}(m)$ is the closest to $a(m)$ in the range
$10<m<20$ by $\alpha =-2.43.$ The ratio $(f_{3}(m) -a(m))/ a(m)$
is shown in Fig. 2 for different values of $A_{1}$ by this value
of $\alpha$.
\begin{figure}[th]
\epsfxsize=8.0cm \centerline{\epsfbox{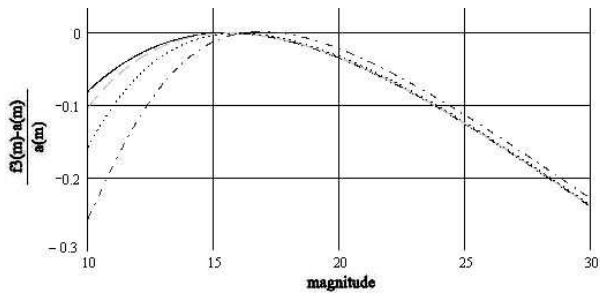}}
\caption{The relative difference $(f_{3}(m)-a(m))/a(m)$ as a
function of the magnitude $m$ for $\alpha=-2.43$ by
$10^{-2}<A_{1}<10^{2}$ (solid), $A_{1}=10^{4}$ (dash),
$A_{1}=10^{5}$ (dot), $A_{1}=10^{6}$ (dadot). }
\end{figure}
If we compare this figure with Figs. 6,10,12 from \cite{77}, we
see that the considered model provides a no-worse fit to
observations than the function $a(m)$ if the same K-corrections
are added for the range $10^2<A_{1}< 10^7$ that corresponds to
$5\cdot 10^{15}>L_{\ast}> 5\cdot 10^{10}.$ {\it Observations
prefer a rising behavior of this ratio up to $m=16,$ and the model
demonstrates it.}
\par
For quasars, I computed the galaxy number counts/redshift relation
$f_{5}(m,z)$ with a different (than for galaxies) luminosity
function $\eta{'} (l(m,z))$ \cite{90}. In Fig 3, there are a
couple of
\begin{figure}[th]
\epsfxsize=8.0cm \centerline{\epsfbox{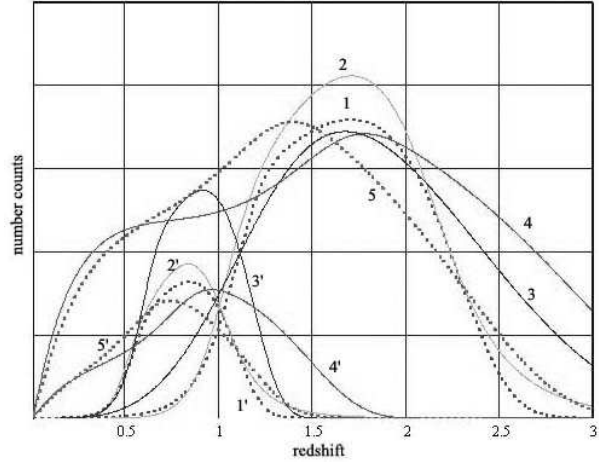}}
\caption{QSO number counts $f_{5}(m,z)$ (arbitrary units) as a
function of the redshift for different luminosity functions:
Gaussian (1', 1 dot), the double power law (2', 2), Schechter's
(3', 3), combined (4', 4 solid and 5', 5 dot) with parameters
given in the text of \cite{90}. The left-shifted curve of each
couple (1' - 5') corresponds to the range $16<m<18.25,$ another
one (1 - 5) corresponds to $18.25<m<20.85.$}
\end{figure}
curves for each case: the left-shifted curve of any couple (1' -
5') corresponds to the range $16<m<18.25,$ another one (1 - 5)
corresponds to $18.25<m<20.85.$ These ranges are chosen the same
as in the paper by Croom et al. \cite{7}, and you may compare this
figure with Fig. 3 in \cite{7}. We can see that the theoretical
distributions reflect only some features of the observed ones but
not an entire picture. Perhaps, it is necessary to consider some
theoretical model of a quasar activity to get a distribution of
"instantaneous" luminosities (a couple of simple examples is
considered in \cite{90}).
\section[5]{Conclusion}
One can verify the quantum and local redshift mechanism of this
model in different ways, but I think that the most cogent one
would be the described prompt measurement of a possible
length-dependent red shift of radiation spectrum in the laboratory
experiment. A negative result of this experiment would be a very
strong support of the standard cosmological model; a positive one
might open the door not only for new cosmology, but for otherwise
quantum gravity, too.

\end{document}